\documentclass[aps,graphicx,twocolumn]{revtex4}
\usepackage{amsmath}
\usepackage{amscd}
\usepackage{graphicx}
\usepackage{multirow}
\usepackage{color}
\begin{document}

\title{Self-assisted complete maximally hyperentangled state analysis via the cross-Kerr nonlinearity}
\author{ Xi-Han Li$^{1,2}$\footnote{
Email address: xihanlicqu@gmail.com}, Shohini Ghose$^{2,3}$}
\address{$^1$ Department of Physics, Chongqing University,
Chongqing, China \\$^2$Department of Physics and Computer Science, Wilfrid Laurier University, Waterloo, Canada\\
$^3$ Institute for Quantum Computing, University of Waterloo, Canada}

\date{\today }
\begin{abstract}
We present two complete maximally hyperentangled state analysis protocols for photons entangled in the polarization and spatial-mode degrees of freedom. The first protocol is a hyperentangled Bell state analysis scheme for two photons and the second is a hyperentangled Greenberger-Horne-Zeilinger (GHZ) state analysis scheme for three photons. In each scheme, a set of mutually orthogonal hyperentangled basis states are completely and deterministically discriminated with the aid of cross-Kerr nonlinearities and linear optics. We also generalize the schemes to unambiguously analyze the $N$-photon hyperentangled GHZ state. Compared with previous protocols, our schemes greatly simplify the discrimination process and reduce the requirements on nonlinearities by using the measured spatial-mode state to assist in the analysis of the polarization state. These advantages make our schemes useful for practical applications in long-distance high capacity quantum communication.

\end{abstract}
\maketitle

\section{Introduction}
Quantum entanglement plays a crucial role in quantum information processing. It is a key resource for quantum communication tasks such as quantum key distribution \cite{qkd1,qkd2}, dense coding \cite{dense1,dense2}, teleportation \cite{tele}, secret sharing \cite{qss1,qss2,qss3}, quantum secure direct communication \cite{qsdc1,qsdc2,qsdc3} and others. Among the many different types of entangled states, the Bell states for two qubits and the Greenberger-Horne-Zeilinger (GHZ) states for three or more qubits are the most popular resources in quantum information processing. These states have two terms in their simplest form and are maximally entangled bases for the Hilbert space. The analysis of Bell states and GHZ states not only has fundamental significance in quantum information theory but is also of practical use in quantum information processing. The number of these basis states that can be distinguished usually determines the capacity of quantum communication schemes. Thus much effort has been made in the past to perform state analysis, with complete discrimination being the ultimate goal. Generally speaking, Bell states and GHZ states can be described by two kinds of information: bit information and phase information. Bit information describes the parities between any two particles and the phase information denotes the relative phase between the two terms. Unambiguous state discrimination can be accomplished by obtaining both kinds of information. In this paper, we present efficient and practical schemes for complete, unambiguous state discrimination of N-photon states that are maximally entangled in both polarization and spatial-mode degrees of freedom. Our schemes take advantage of information about the spatial-mode state to efficiently analyze the polarization state of the photons, thus significantly improving on past protocols.

Although a set of mutually orthogonal basis states can in theory be completely discriminated, complete Bell state analysis (BSA) and GHZ state analysis (GSA) of photons cannot be realized by linear optics alone, without resorting to ancillaries \cite{bsa1,bsa2,bsa3,gsa}. It has been shown that the four Bell states for two photons can only be classified into three groups and the optimal success probability of state analysis is 50$\%$ using only linear optics. However, the success probability can be improved by several means, for example, the use of auxiliary entanglement in another degree of freedom (DOF)\cite{CBSAH1,CBSAH2,CBSAH3,CBSAH4,CBSAH5}, via assistant states \cite{BSAS, BSAS1} and through a nonlinear interaction such as the cross-Kerr nonlinearity \cite{BSAN}.

Whereas traditional entanglement involves particles that are only entangled in one degree of freedom, hyperentanglement involves particles simultaneously entangled in more than one degree of freedom, and has attracted much attention in recent years. There are several DOFs of a photon such as polarization, spatial mode, time-bin, frequency, etc, that can be used to construct hyperentangled states. Hyperentanglement has the appealing feature that each photon carries information in two or more DOFs and the DOFs can be manipulated independently. This can improve both the security and the channel capacity of quantum communications \cite{ah1}. Recently, experimental preparation of hyperentanglement has been reported \cite{preh1,preh2,preh3}. Hyperentanglement has many applications in quantum information processing, such as complete Bell-state analysis \cite{CBSAH1,CBSAH2,CBSAH3,CBSAH4,CBSAH5}, hyper-parallel quantum computing \cite{ah2,ah3}, deterministic entanglement purification protocols \cite{ah4,ah5,ah6,ah7} and quantum repeaters \cite{ah8}. There has also been interesting recent progress in hyperentanglement concentration and hyperentanglement purification~\cite{hchp1,hchp2,hchp3,hchp4,hchp5,hchp6,hchp7,hchp8,hchp9}.

To date, the most popular hyperentangled states of photons are those entangled in polarization and spatial-mode DOFs, since the manipulation techniques for these two DOFs is mature. In high-capacity quantum communication schemes that utilize hyperentangled channels, hyperentangled Bell state analysis (HBSA) and hyperentangled GHZ state analysis (HGSA) are key steps required to read the information.  Hyperentangled state discrimination needs to confirm the bit and phase information for both DOFs, which is more challenging than traditional entangled state analysis. Considering both DOFs together, there are 16 hyperentangled Bell states in total for two photons. It has been shown that these 16 states can be classified into only 7 groups via linear optics \cite{HBSA1,HBSA2} and thus cannot be completely distinguished. In 2010, Sheng \emph{et al}. proposed the first complete HBSA scheme using the cross-Kerr nonlinearity \cite{CHBSA1}. In their scheme, three quantum nondemolition detectors (QNDs) constructed using cross-Kerr nonlinearities are used to read out the bit and phase information of the spatial-mode state and the bit information of the polarization state respectively. Then the phase information of the polarization DOF is obtained by measurement in the diagonal basis of the polarization state. With these 4 bits of information, 16 polarization-momentum hyperentangled Bell states can be unambiguously discriminated. Later, Xia \emph{et al}. presented an efficient HGSA protocol using a similar principle \cite{CHGSA}. In these two schemes, the spatial-mode state is analyzed in the first step. Although the QNDs preserve the photons , the coherence of the spatial-mode state is destroyed. If preserved, it can be useful for the discrimination of the polarization state. Unlike complete BSA and GSA schemes which resort to several auxiliary tools such as additional entanglement, ancillary states or nonlinear interactions, the main resource for complete HBSA and HGSA is the nonlinearity. Complete HBSA schemes were also realized with the help of quantum-dot spins in optical microcavities and nitrogen-vacancy centers in resonators \cite{CHBSA2,CHBSA3,CHBSA4}. Recently, Liu\emph{ et al.} proposed a complete nondestructive analysis of the two-photon six-qubit hyperentangled Bell states assisted by cross-Kerr nonlinearity, in which the photons are entangled simultaneously in the polarization and two longitudinal momentum DOFs \cite{CHBSAT}.

In this paper we first present a simplified complete HBSA scheme which deterministically distinguishes 16 hyperentangled Bell states of two photons. The bit and phase information of the spatial-mode state is read by two QNDs constructed with the cross-Kerr nonlinearity \cite{kerr}. The key point is that neither the photons nor the spatial-mode state will be destroyed by the QNDs. With the help of the preserved spatial-mode entanglement, the bit and phase information of the polarization state can be deduced simultaneously by two single-photon Bell state measurements (SPBSMs). We then describe a complete HGSA protocol for the three-photon hyperentangled GHZ state, in which three assistant coherent states and three SPBSMs are required. With our scheme, 64 three-photon hyperentangled GHZ states can be unambiguously discriminated. By maintaining the spatial-mode coherence in the first step and using it to assist in the discrimination of polarization states, our schemes greatly simplify the process and significantly reduce the required  nonlinearities compared with previous protocols. We also generalize the scheme to a complete and deterministic analysis of $N$-photon hyperentangled GHZ states. A detailed discussion and summary is provided in the last section.

\section{Complete hyperentangled Bell state analysis}
The two-photon hyperentangled Bell state can be written as
\begin{eqnarray}
  \vert \Upsilon \rangle_{AB} = \vert \Theta_P\rangle_{AB}\otimes \vert\Xi_S\rangle_{AB}.
\end{eqnarray}
Here $A$ and $B$ denote the two photons and the subscripts $P$ and $S$ represent the polarization and spatial-mode DOF, respectively.
$\vert \Theta_P\rangle_{AB}$ can be one of the following four Bell states in the polarization DOF,
\begin{eqnarray}
  \vert \Phi^{\pm}\rangle_P&=&\frac{1}{\sqrt{2}} (\vert HH\rangle \pm \vert VV \rangle)_{AB}, \\
  \vert \Psi^{\pm}\rangle_P&=&\frac{1}{\sqrt{2}} (\vert HV\rangle \pm \vert VH \rangle)_{AB}.
\end{eqnarray}
$\vert H\rangle$ and $\vert V\rangle$ indicate the horizontal and the vertical polarizations, respectively. The spatial-mode state $\vert\Xi_S\rangle_{AB}$ is one of the four Bell states in the spatial modes
\begin{eqnarray}
   \vert \Phi^{\pm}\rangle_S&=&\frac{1}{\sqrt{2}} (\vert a_1b_1\rangle \pm \vert a_2b_2 \rangle)_{AB}, \\
  \vert \Psi^{\pm}\rangle_S&=&\frac{1}{\sqrt{2}} (\vert a_1b_2\rangle \pm \vert a_2b_1 \rangle)_{AB}.
\end{eqnarray}
Here $a_1 (b_1)$ and $a_2 (b_2)$ are the two possible spatial modes of photon $A (B)$. Taking into account the two DOFs together, there are 16 hyperentangled Bell states and our task is to distinguish them completely.

Before we describe our scheme, we introduce the basic principle of the cross-Kerr nonlinearities which play a central role in our spatial-mode state discrimination. The Hamiltonian describing the interaction between a signal state $\vert \psi\rangle_s$ and a probe coherent state $\vert \alpha \rangle_p$ in the nonlinear medium can be written as
\begin{eqnarray}
  H=\hbar\chi a^\dagger_s a_s a_p^\dagger a_p.
\end{eqnarray}
Here $a_s^\dagger (a_p^\dagger)$ and $a_s (a_p)$ are the creation and annihilation operations for the signal (probe) state, respectively. $\chi$ is the coupling strength of the nonlinearity and depends on the material. After the interaction with the signal state in the medium, the coherent state picks up a phase shift which is proportional to the photon number $N$ of the signal state,
\begin{eqnarray}
  \vert \alpha\rangle_p \rightarrow \vert \alpha e^{iN\theta}\rangle.
\end{eqnarray}
Here $\theta=\chi t$ and $t$ is the interaction time. By measuring the phase shift via the $X$-quadrature measurement, the number of photons can be read out without destroying the photons. We can choose $X$-quadrature measurements that do not distinguish phase shifts differing in sign ``$\pm$". This feature preserves the coherence of photons with respect to each other as well as the photons themselves. The cross-Kerr nonlinearity has been widely used to construct quantum nondemolition detectors (QNDs) for quantum information processing in the past decade \cite{BSAN,ah4,hchp7,CHBSA1,CHGSA,kerr,kerr0}. In our scheme, it is used in two QNDs which read the bit and phase information of the spatial-mode state.

\begin{figure}[!h]
\centering
\includegraphics*[width=3.2in]{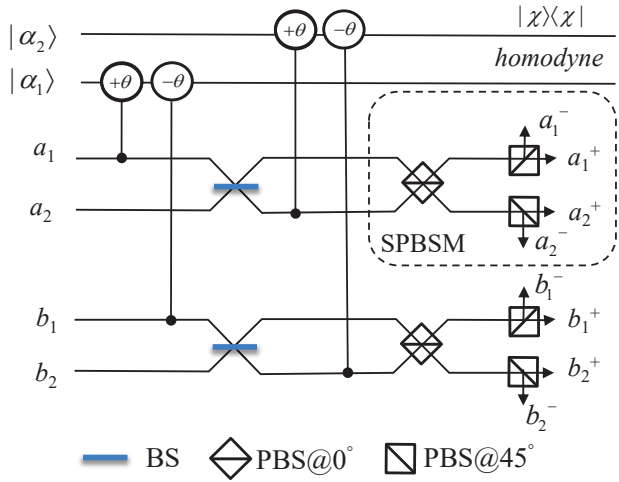}
\caption{Schematic diagram of our complete HBSA protocol. The four cross-Kerr nonlinear interactions produce a phase shift of $\pm \theta$ on the coherent states $\vert \alpha_1 \rangle$ and $\vert \alpha_2 \rangle$ if photons are in corresponding spatial modes. The beam splitters (BSs) guide the photons from each input port to these two output ports with equal probabilities. The polarizing beam splitters (PBSs) at $0^\circ$ transmit horizontal polarized states while reflecting vertical polarized states. The PBSs at $45^\circ$ transmit $\vert +\rangle=\frac{1}{\sqrt{2}}(\vert H\rangle+\vert V\rangle)$, and reflect $\vert -\rangle=\frac{1}{\sqrt{2}}(\vert H\rangle-\vert V\rangle)$. The dashed rectangle represents a single-photon Bell state measurement (SPBSM), which discriminates four single-photon Bell states completely. With the homodyne measurements on the two coherent states and two SPBSMs, 16 hyperentangled Bell states can be completely distinguished. }
\end{figure}

The setup of our proposed HBSA protocol is shown in Fig.1. The process consists of two steps: discrimination of the spatial-mode Bell states via the cross-Kerr nonlinearity, followed by discrimination of polarization Bell states assisted by the spatial-mode entanglement. We now introduce the process step by step.

After the photons in $a_1$ and $b_1$ interact with the coherent state $\vert \alpha_1\rangle$, the state of the collective system evolves as
\begin{eqnarray}
  \vert \Phi^{\pm}\rangle_S \vert \alpha_1\rangle &=& \frac{1}{\sqrt{2}}(\vert a_1b_1\rangle \pm \vert a_2b_2 \rangle)\vert \alpha_1\rangle\nonumber\\
  &\rightarrow& \frac{1}{\sqrt{2}}(\vert a_1b_1\rangle \pm \vert a_2b_2 \rangle)\vert \alpha_1\rangle= \vert \Phi^{\pm}\rangle_S \vert \alpha_1\rangle, \\
  \vert \Psi^{\pm}\rangle_S \vert \alpha_1\rangle&=&\frac{1}{\sqrt{2}} (\vert a_1b_2\rangle \pm \vert a_2b_1 \rangle)\vert \alpha_1\rangle\nonumber\\
   &\rightarrow& \frac{1}{\sqrt{2}} (\vert a_1b_2\rangle\vert \alpha_1e^{i\theta}\rangle \pm \vert a_2b_1 \rangle\vert \alpha_1e^{-i\theta}\rangle)\nonumber\\
   &=& \vert \Psi^{\pm}\rangle_S \vert \alpha_1e^{\pm i\theta}\rangle.
\end{eqnarray}
Here we omit the polarization DOF since it is invariant during the evolution.  Note that the final line in (9) follows from the fact that the $X$-quadrature measurement on $\vert \alpha_1\rangle$ is set up to only distinguish the phase 0 from $\pm \theta$. Hence, with this measurement, $\vert \Phi^{\pm}\rangle_S$ can be discriminated from $\vert \Psi^{\pm}\rangle_S$. In other words, the bit information about the spatial-mode is obtained.

The two spatial modes of each photon are then mixed at the beam splitters (BSs), which act as a Hadamard operation on the spatial-mode DOF,
\begin{eqnarray}
  \vert x_1\rangle &\rightarrow& \frac{1}{\sqrt{2}}(\vert x_1\rangle+\vert x_2\rangle), \\
  \vert x_2\rangle &\rightarrow& \frac{1}{\sqrt{2}}(\vert x_1\rangle-\vert x_2\rangle).
\end{eqnarray}
Here $x$ denotes $a$ or $b$. The effect of the two BSs is to transform the input spatial-mode Bell states to different spatial-mode Bell states. In detail, $\vert \Phi^+\rangle_S$ and $\vert \Psi^+\rangle_S$ are invariant, while $\vert \Phi^-\rangle_S \rightleftharpoons \vert \Psi^+\rangle_S$. 
The form of the nonlinear interactions between the photons and the second coherent state $\vert \alpha_2\rangle$ is similar to the first one, which distinguishes the original $\vert \Phi^{+}\rangle_S (\vert \Psi^{+}\rangle_S)$ from $\vert \Phi^{-}\rangle_S (\vert \Psi^{-}\rangle_S)$. This provides the phase information, and hence the four spatial-mode Bell states are completely discriminated. Meanwhile, the spatial-mode entanglement is also changed. The relations between the original spatial-mode Bell state, the new spatial-mode Bell state and the phase shifts of the two coherent beams are shown in Table. I.
\begin{table}[!h]
\caption{Corresponding relations between the original state, the new state after the BS's in Fig 1, and the two phase shifts of coherent states. }
\begin{tabular}{c|c|c|c}
\hline\hline
Original state & New state & $\vert \alpha_1\rangle$ & $\vert \alpha_2\rangle$ \\
\hline
$\vert \Phi^{+}\rangle_S$ & $\vert \Phi^{+}\rangle_S$ & 0 & 0\\
$\vert \Phi^{-}\rangle_S$ & $\vert \Psi^{+}\rangle_S$ & 0 & $\pm\theta$\\
$\vert \Psi^{+}\rangle_S$ & $\vert \Phi^{-}\rangle_S$ & $\pm\theta$ & 0\\
$\vert \Psi^{-}\rangle_S$ & $\vert \Psi^{-}\rangle_S$ & $\pm\theta$ & $\pm\theta$\\
\hline
\end{tabular}
\end{table}

In the second step, two single-photon Bell state measurements (SPBSMs) are performed on the two photons, whose measurement outcomes will result in the complete discrimination of four polarization Bell states. The four single-photon Bell states composed of the polarization and spatial-mode DOFs are
\begin{eqnarray}
  \vert \phi^{\pm}\rangle_X &=& \frac{1}{\sqrt{2}}(\vert Hx_2\rangle \pm\vert Vx_1\rangle)_X, \\
  \vert \psi^{\pm}\rangle_X &=& \frac{1}{\sqrt{2}}(\vert Hx_1\rangle \pm\vert Vx_2\rangle)_X.
\end{eqnarray}
Here $X(x)$ can be either $A(a)$ or $B(b)$. After the polarizing beam splitters (PBSs) at 0$^\circ$ (which transmit horizontal states while reflecting vertical ones) and PBSs at 45$^\circ$ (which transmit $\vert +\rangle=\frac{1}{\sqrt{2}}(\vert H\rangle+\vert V\rangle)$ state and reflect $\vert -\rangle=\frac{1}{\sqrt{2}}(\vert H\rangle-\vert V\rangle)$), four single-photon Bell states will trigger the four different detectors placed in the four output ports, accordingly. Specifically, $\vert \phi^{\pm}\rangle_X$ goes to $x^\pm_1$ while $\vert \psi^{\pm}\rangle_X$ goes to $x^\pm_2$.

In this step, the spatial-mode state is known and provides important assistance in the analysis of the four polarization Bell states. For example, if the new spatial-mode state after the first step is $\vert \Psi^{-}\rangle_S$, the four possible corresponding hyperentangled states will result in different combinations of SPBSMs as
\begin{eqnarray}
\vert \Phi^{\pm}\rangle_P\otimes \vert \Psi^{-}\rangle_S &=& \frac{1}{2}(\vert \psi^\pm\rangle_A\vert \phi^-\rangle_B+\vert \psi^\mp\rangle_A\vert \phi^+\rangle_B\nonumber\\
  &&-\vert \phi^\pm\rangle_A\vert \psi^-\rangle_B-\vert \phi^\mp\rangle_A\vert \psi^+\rangle_B),\\
\vert \Psi^{\pm}\rangle_P\otimes \vert \Psi^{-}\rangle_S &=& \frac{1}{2}(\vert \psi^-\rangle_A\vert \psi^\pm\rangle_B-\vert \psi^+\rangle_A\vert \psi^\mp\rangle_B\nonumber\\
  &&-\vert \phi^+\rangle_A\vert \phi^\mp\rangle_B-\vert \phi^-\rangle_A\vert \phi^\pm\rangle_B).
\end{eqnarray}
There are 16 possible measurement combinations, which can be collected into four groups.
Each group corresponds to a specific polarization Bell state.
Therefore, all four possible polarization Bell states can be deterministically discriminated by identifying which group the measurement outcomes belong to. If the spatial-mode state is one of the other three Bell states, the four polarization states can also be distinguished in the same way. The detailed relations are shown in Table. II. For example, if the two SPBSM results are $\vert \psi^-\rangle_A$ and $\vert \phi^+\rangle_B$, the new state after the first step belongs to the last group. If the first step determines that the spatial-mode state is $\vert \Phi^+\rangle_S$, one can deduce that the polarization state is $\vert \Psi^-\rangle_P$.

 \begin{table}[!h]
\caption{Relations between the new state before the second step in Fig. 1 and possible detections. }
\begin{tabular}{c|cl}
\hline\hline
New states & Possible detections  \\
\hline
$\vert \Phi^{+}\rangle_P\otimes \vert \Phi^{+}\rangle_S$, $\vert \Phi^{-}\rangle_P\otimes \vert \Phi^{-}\rangle_S$,& $\vert \phi^{+}\rangle_A\vert \phi^{+}\rangle_B$, $\vert \phi^{-}\rangle_A\vert \phi^{-}\rangle_B$, \\$\vert \Psi^{+}\rangle_P\otimes \vert \Psi^{+}\rangle_S$,$\vert \Psi^{-}\rangle_P\otimes \vert \Psi^{-}\rangle_S$. & $\vert \psi^{+}\rangle_A\vert \psi^{+}\rangle_B$, $\vert \psi^{-}\rangle_A\vert \psi^{-}\rangle_B$. \\
\hline
$\vert \Phi^{+}\rangle_P\otimes \vert \Psi^{+}\rangle_S$, $\vert \Phi^{-}\rangle_P\otimes \vert \Psi^{-}\rangle_S$, & $\vert \phi^{+}\rangle_A\vert \psi^{+}\rangle_B$, $\vert \phi^{-}\rangle_A\vert \psi^{-}\rangle_B$, \\ $\vert \Psi^{+}\rangle_P\otimes \vert \Phi^{+}\rangle_S$, $\vert \Psi^{-}\rangle_P\otimes \vert \Phi^{-}\rangle_S$. & $\vert \psi^{+}\rangle_A\vert \phi^{+}\rangle_B$, $\vert \psi^{-}\rangle_A\vert \phi^{-}\rangle_B$. \\
\hline
$\vert \Phi^{+}\rangle_P\otimes \vert \Phi^{-}\rangle_S$, $\vert \Phi^{-}\rangle_P\otimes \vert \Phi^{+}\rangle_S$,& $\vert \phi^{+}\rangle_A\vert \phi^{-}\rangle_B$, $\vert \phi^{-}\rangle_A\vert \phi^{+}\rangle_B$,\\ $\vert \Psi^{+}\rangle_P\otimes \vert \Psi^{-}\rangle_S$, $\vert \Psi^{-}\rangle_P\otimes \vert \Psi^{+}\rangle_S$. & $\vert \psi^{+}\rangle_A\vert \psi^{-}\rangle_B$, $\vert \psi^{-}\rangle_A\vert \psi^{+}\rangle_B$. \\
\hline
$\vert \Phi^{+}\rangle_P\otimes \vert \Psi^{-}\rangle_S$, $\vert \Phi^{-}\rangle_P\otimes \vert \Psi^{+}\rangle_S$,& $\vert \phi^{+}\rangle_A\vert \psi^{-}\rangle_B$, $\vert \phi^{-}\rangle_A\vert \psi^{+}\rangle_B$,\\$\vert \Psi^{+}\rangle_P\otimes \vert \Phi^{-}\rangle_S$, $\vert \Psi^{-}\rangle_P\otimes \vert \Phi^{+}\rangle_S$. & $\vert \psi^{+}\rangle_A\vert \phi^{-}\rangle_B$, $\vert \psi^{-}\rangle_A\vert \phi^{+}\rangle_B$. \\
\hline
\end{tabular}
\end{table}
From the preceding analysis, the 16 hyperentangled Bell states are completely discriminated with our two-step scheme. The distinguishing of the polarization state is aided by spatial-mode entanglement. We note that although our scheme consists of two steps, there is no need to pause the state analysis procedure midway. The information about the spatial-mode state can be used to deduce the polarization state after all the measurements have been performed.

\section{Complete hyperentangled Greenberger-Horne-Zeilinger state analysis}
In this section, we introduce the complete HGSA scheme by describing the three-photon hyperentangled GHZ state as an example first. Then we generalize the scheme to analyze the $N$-photon hyperentangled GHZ state.

Generally, the $N$-photon hyperentangled GHZ state in both polarization and spatial-mode DOFs can be written as
\begin{eqnarray}
  \vert \Upsilon\rangle_{AB...Z}=\vert \Theta_P\rangle_{AB...Z}\otimes \vert \Xi_S\rangle_{AB...Z}.
\end{eqnarray}
Here $A$, $B$,...,$Z$ denote the $N$ photons. There are $2^N$ maximally entangled GHZ states in each DOF, which can be written in a unified form as
\begin{eqnarray}
  \vert \Omega^\pm_{\texttt{ab...z}} \rangle_{AB...Z} &=& \frac{1}{\sqrt{2}}(\vert \texttt{ab...z}\rangle\pm \vert \bar{\texttt{a}}\bar{\texttt{b}}...\bar{\texttt{z}}\rangle)_{AB...Z}.
\end{eqnarray}
 Here $\texttt{ab...z}\in\{0,1\}$ refer to the bit information and $\bar{\texttt{x}}=1-\texttt{x}, (\texttt{x}=\texttt{ab...z})$. For the polarization DOF, $\vert 0\rangle\equiv \vert H\rangle$, $\vert 1\rangle\equiv \vert V\rangle$. For the spatial-mode DOF, $\vert 0\rangle\equiv \vert x_1\rangle$ and $\vert 1\rangle\equiv \vert x_2\rangle$ $x=a,b,...z$. Since $\vert \Omega^\pm_{\texttt{ab...z}}\rangle$ and $\vert \Omega^\pm_{\bar{\texttt{a}}\bar{\texttt{b}}...\bar{\texttt{z}}}\rangle$ only differ by a nonessential global phase, we limit the number of "1"s in the subscript string ``$\texttt{ab...z}$" of $\Omega$ to be no larger than $N/2$ to ensure that there are $2^N$ mutually orthogonal entangled GHZ states in total. (When N is even, we choose the subscript string with a lower binary value.)

We discuss the $N=3$ situation first. One of these 64 hyperentangled GHZ states for example is
 \begin{eqnarray}
   &&\vert \Omega^+_{000}\rangle_P\otimes \vert \Omega^+_{000}\rangle_S \nonumber\\
   &=& \frac{1}{\sqrt{2}}(\vert HHH\rangle+\vert VVV\rangle)\otimes \frac{1}{\sqrt{2}}(\vert a_1b_1c_1\rangle+\vert a_2b_2c_2\rangle).\nonumber\\
 \end{eqnarray}
The set-up of our complete HGSA scheme for three-photon states is shown in Fig.2. In this scheme, three coherent states are employed to distinguish the eight spatial-mode GHZ states. The first two are used to check the parity between $AB$ and $AC$, respectively, i.e., read out the bit information. The third one is used to confirm the relative phase information ``$\pm$". If the number of $\pi$ phase shifts is odd, which results in a single overall $\pi$ phase shift, the relative phase information of the spatial-mode state is ``$-$". Otherwise, an even number of $\pi$ shifts leads to zero phase shift, which indicates the ``$+$" relative phase information for the spatial-mode. For example, 
\begin{eqnarray}
 && \vert \Omega^+_{000}\rangle_S \vert \alpha_3\rangle \nonumber\\&\rightarrow& \frac{1}{2}(\vert a_1b_1c_1\rangle+\vert a_2b_2c_1\rangle+\vert a_2b_1c_2\rangle+\vert a_1b_2c_2\rangle)\vert \alpha_3\rangle,\\
 && \vert \Omega^-_{000}\rangle_S \vert \alpha_3\rangle \nonumber\\&\rightarrow& \frac{1}{2}(\vert a_2b_2c_2\rangle+\vert a_2b_1c_1\rangle+\vert a_1b_2c_1\rangle+\vert a_1b_1c_2\rangle)\vert \alpha_3e^{i\pi}\rangle.\nonumber\\
\end{eqnarray}

The relations between the original state and these three measured phase shifts are shown in Table. III. Then three BSs are used to manipulate the spatial-mode state back to its initial status. In the second step, three SPBSMs are performed. Based on the SPBSM results and the information about the spatial-mode state in step one, the eight polarization GHZ states can be completely discriminated and consequently the 64 hyperentangled GHZ states can be distinguished.

\begin{figure}[!h]
\centering
\includegraphics*[width=3.2in]{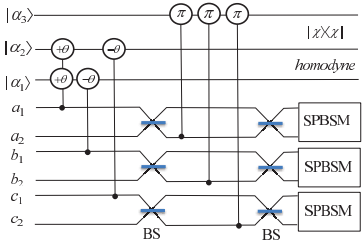}
\caption{Schematic diagram of our complete HGSA protocol for three-photon hyperentangled GHZ states. Three coherent states are utilized to distinguish the spatial-mode GHZ states. The circles represent nonlinear interactions and the phases inside the circles are the phase shift of the coherent states. SPBSM denotes a single-photon Bell state measurement as shown in Fig.1, which can discriminate four single-photon Bell states with certainty.}
\end{figure}

\begin{table}[!h]
\caption{Corresponding relations between the original state and the three phase shifts of the coherent states in Fig. 2. }
\begin{tabular}{c|c|c|c}
\hline\hline
Original state &  $\vert \alpha_1\rangle$ & $\vert \alpha_2\rangle$ &$\vert \alpha_3\rangle$\\
\hline
$\vert \Phi^{+}_{000}\rangle_S$ &  0 & 0 & 0 \\
$\vert \Phi^{-}_{000}\rangle_S$ &  0 & 0 & $\pi$ \\
$\vert \Phi^{+}_{001}\rangle_S$ &  0 & $\pm \theta$ & 0 \\
$\vert \Phi^{-}_{001}\rangle_S$ &  0 & $\pm \theta$ & $\pi$ \\
$\vert \Phi^{+}_{010}\rangle_S$ & $\pm \theta$ & 0 & 0 \\
$\vert \Phi^{-}_{010}\rangle_S$ &  $\pm \theta$ & 0 & $\pi$ \\
$\vert \Phi^{+}_{100}\rangle_S$ & $\pm \theta$ & $\pm \theta$ & 0 \\
$\vert \Phi^{-}_{100}\rangle_S$ &  $\pm \theta$ & $\pm \theta$ & $\pi$ \\
\hline
\end{tabular}
\end{table}

We denote $\varphi$ to be the phase information of a state
\begin{eqnarray}
  \varphi_{\vert \Omega^\pm\rangle} \equiv\pm,  \varphi_{\vert \phi^\pm\rangle} \equiv\pm,  \varphi_{\vert \psi^\pm\rangle}\equiv\pm.
\end{eqnarray}
We also define the bit information of single-photon Bell states as
\begin{eqnarray}
  \delta_{\vert \psi^\pm\rangle} \equiv 1 , \delta_{\vert \phi^\pm\rangle} \equiv 0.
\end{eqnarray}
The corresponding relations between the original hyperentangled GHZ state and the SPBSM results are
\begin{eqnarray}
  &&\varphi_P\otimes \varphi_S=\varphi_A\otimes \varphi_B \otimes \varphi_C, \\
  &&\texttt{x}_P\oplus \texttt{x}_S= \delta_X. (\texttt{x=a,b,c}, X=A,B,C.)
\end{eqnarray}
With these relations, the polarization state can be deduced. For example, if the three single-photon Bell states are $\vert \psi^+\rangle_A\vert \psi^-\rangle_B\vert \phi^+\rangle_C$ and the spatial-mode state is $\vert \Omega^+_{010}\rangle_S$, we have
\begin{eqnarray}
  &&\varphi_P=\varphi_A\otimes \varphi_B \otimes \varphi_C\otimes \varphi_S=-,\\
  &&\delta_A\delta_B\delta_C= 110,\\
  &&\texttt{a}_P=\delta_A\oplus \texttt{a}_S=1, \\
  &&\texttt{b}_P=\delta_B\oplus \texttt{b}_S=0, \\
  &&\texttt{c}_P=\delta_C\oplus \texttt{c}_S=0.
\end{eqnarray}
The answer is $\vert \Omega^-_{100}\rangle_P$ and the original hyperentangled GHZ state is $\vert \Omega^-_{100}\rangle_P \otimes \vert \Omega^+_{010}\rangle_S$.
Our scheme can distinguish the 64 hyperentangled GHZ states deterministically.

It is possible to generalize our scheme to analyze $N$-photon hyperentangled GHZ states as shown in Fig.3. Firstly, $N-1$ parity checks are performed on photon pairs $AB$, $AC$,... $AZ$ with the help of $N-1$ auxiliary coherent states $\vert \alpha_n\rangle (n=1,2,...N-1)$ to read the bit information of the spatial-mode states. After $X$-quadrature measurements on these $N-1$ coherent states, $2^N$ spatial-mode GHZ states can be placed into $2^{N-1}$ groups, and the parity information ``$\texttt{a}_S\texttt{b}_S...\texttt{z}_S$" can be identified. Then after the effect of $N$ BSs and interactions between photons and a coherent state $\vert \alpha_N\rangle$, the relative phase information ``$\pm$" of the states in each group can be further detected. The polarization GHZ state can be deduced based on the spatial-mode state information and $N$ SPBSMs following a similar procedure as described in the preceding three-photon HGSA scheme. Thus the $4^N$ hyperentangled GHZ states can be completely distinguished.

\begin{figure}[!h]
\centering
\includegraphics*[width=3.2in]{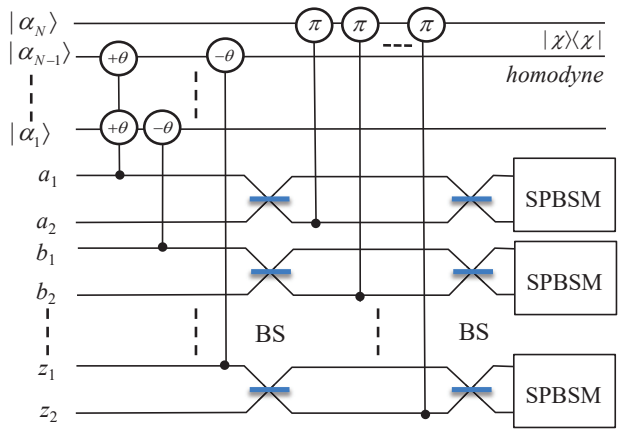}
\caption{Schematic diagram of our complete HGSA protocol for $N$-photon hyperentangled GHZ states. $N$ coherent states are used to discriminate the spatial-mode states without destroying the entanglement. Then $N$ SPBSMs are used to measure the polarization state with the help of the information about the spatial-mode state.  }
\end{figure}

\section{Discussion and summary}
This paper presents two complete hyperentangled state analysis schemes and gives a general protocol for distinguishing $N$-photon dual hyperentanglement in polarization and spatial-mode states. Using our schemes, entangled states in both DOFs can be unambiguously discriminated. In our protocols, the spatial-mode states are analyzed by QNDs constructed using cross-Kerr nonlinearities, which is a challenging task  with current technology. However, although the natural cross-Kerr nonlinearities are weak and the Kerr phase shift is small at the single-photon level, recent research shows promising progress towards practical use of the effect in the near future. The magnitude of $\theta\approx10^{-18}$ of the natural cross-Kerr nonlinearities can be improved to magnitude $\sim10^{-2}$ by electromagnetically induced transparencies and other means \cite{BSAN}. In 2003, Hofmann \emph{et al.} showed that a phase shift of $\pi$ can be achieved with a single two-level atom in a one-sided cavity \cite{kerr1}. In 2011, Feizpour
\emph{et al} showed that an observable value amplified from a single-photon-level cross-Kerr phase shift by using weak-value amplification is possible \cite{kerr2}. Later, a device that can amplify the nonlinearity effect was also proposed  to construct a two-qubit parity gate with tiny cross-Kerr nonlinearity\cite{kerr2+}. In 2013, a giant cross-Kerr effect induced by an artificial atom was reported in which average cross-Kerr phase shifts of up to 20 degrees per photon with coherent microwave fields at the single-photon level were demonstrated \cite{kerr3}. Moreover, Ref. \cite{kerr4} has shown that giant cross Kerr nonlinearities of the probe and the signal pulses can be obtained with nearly vanishing optical absorption, based on which two-qubit quantum polarization phase gates can be constructed. In addition, to enhance the nonlinearities, improvement of the measurement on the coherent state can also promote the feasibility of the nonlinearity. In 2010, Wittmann \emph{et al} showed the displacement-controlled photon number resolving detector surpassed the standard homodyne detector \cite{kerr5}. 
It is fortunate that our complete HBSA scheme only requires a small phase shift, and the scheme will succeed as long as the small phase shift can be distinguished from zero. For a weak cross-Kerr nonlinearity, a sufficiently large amplitude of the coherent state that satisfies $\alpha \theta^2>>1$ can make it possible to distinguish a small phase shift in the coherent state from 0 phase shift. In Ref.\cite{BSAN}, the estimated value of $\alpha^2$ is $1.3\times 10^4$ provided that a cryogenic NV-diamond system can generate a phase shift of more than 0.1 rad per signal photon. In 2009, the first experimental observation of optical-fiber Kerr nonlinearity at the single-photon level was demonstrated \cite{kerr6}. Most recently, an implementation of strong optical nonlinearity using electromagnetically induced transparency was presented and a nonlinear phase shift was measured \cite{kerr7}.
All these studies indicate that our HBSA scheme is feasible with current technology. On the other hand, our complete HGSA protocol requires nonlinearities that can generate a $\pi$ phase shift. Although it is not easy to realize with current technology, it is necessary in our scheme for reading the phase information of the spatial-mode state without destroying its entanglement. In this paper, we have demonstrated the principle by using the cross-Kerr nonlinearity as an example. There are many other kinds of interaction which can also provide feasible ways to realize the function we need \cite{non1,non2,non3,non4,non5}.

Hyperentanglement analysis plays an important role in quantum information processing based on hyperentangled states. Typical applications are hyperentanglement swapping, hyperdense coding and teleportation via hyperentangled channels \cite{CHBSA1,CHBSA2,CHBSA4,CHBSAT}. Moreover it is also useful in establishing quantum repeaters and high-capacity quantum communication. Our schemes simplify the analysis process and reduce the resources, which will make these applications more feasible and economical.

In our protocols, the spatial-mode state is discriminated first followed by the polarization state. In theory, the two DOFs are equivalent and the order can be reversed. However, distinguishing polarization states first will consume more resources due to the uncertainty of spatial modes.
It is interesting to compare our schemes with the previously proposed complete HBSA \cite{CHBSA1} and complete HGSA \cite{CHGSA} schemes. In both these two schemes, the bit and phase information of the spatial-mode state and the bit information of the polarization state are read out using QNDs. After the analysis of the spatial-mode state, the coherence of the spatial-mode state is no longer maintained, i.e., the state collapses to a product state. Therefore, the process should be paused to confirm the photons' spatial modes before discrimination of the polarization states. Otherwise, more QNDs should be prepared in advance for all possible spatial-mode states before reading out the bit information of the polarization state. However, in our schemes, the spatial-mode entanglement is kept intact after its analysis and is subsequently used to assist in the discrimination of the polarization state. Compared with previous schemes, our protocols thus have some distinctive features: (i) There is no requirement to pause the process since our schemes can be implemented in one-shot. (ii) The polarization state is discriminated without resorting to any nonlinearities, which greatly reduces the requirement on nonlinear interactions.  (iii) Although QNDs are used to read out the phase information of spatial-mode states, which looks similar to the previous protocols, only two potential phase shifts need to be distinguished in our schemes as opposed to four phase shifts in both the previous HBSA and HGSA protocols. In conclusion, our schemes reduce the number of nonlinearities required and save time and quantum resources, making them simple and feasible in practical applications.
%


\section*{Acknowledgement}

XL is supported by the National Natural Science
Foundation of China under Grant Nos. 11574038 and 11547305. SG acknowledges support from the Natural Sciences and Engineering Research Council of Canada.

\end{document}